\documentclass[aps,showpacs,superscriptadress,preprint]{revtex4}
\usepackage{amssymb}
\usepackage{graphicx}
\begin{document}
\title{Temperature effect on the power spectrum in inflation}

\author{Shaoyu Yin$^1$\footnote{051019008@fudan.edu.cn}, Bin Wang$^1$\footnote{wangb@fudan.edu.cn},
Ru-Keng Su$^{1,2}$\footnote{rksu@fudan.ac.cn}} \affiliation{
\small 1. Department of Physics, Fudan University,Shanghai 200433, China\\
\small 2. CCAST(World Laboratory), P.O.Box 8730, Beijing 100080,
China}

\begin{abstract}
We examine the effect of the thermal vacuum on the power spectrum of
inflation by using the thermal field dynamics. We find that the
thermal effect influences the CMB anisotropy at large length scale.
After removing the divergence by using the holographic cutoff, we
observe that the thermal vacuum explains well the observational CMB
result at low multipoles. This shows that the temperature dependent
factor should be considered in the study of power spectrum in
inflation, especially at large length scale.
\end{abstract}


\maketitle

The scenario of the inflation provides an excellent solution to the
horizon and flatness problems \cite{Guth:1981}. During inflation the
quantum fluctuations in the inflaton field can freeze out and become
the seed for large scale structure \cite{Linde:1990,Liddle:2000}.
These fluctuations can translate into perturbations in energy
density and curvature in the universe and can be imprinted in the
anisotropy of cosmic microwave background (CMB), which can be
measured with higher and higher precision nowadays. Recently an
intriguing possibility has been discovered in the literature that
inflation might provide a window towards physics beyond the Planck
scale \cite{Easther:2001,Easther:2002,Danielsson:2002a,Martin:2004,
Martin:2005&2006,Groeneboom:2008,Ren:2006,Wang:2003,Goldstein:2003}.
Chances to detect such trans-Planckian effect in the CMB observation
have been addressed in Refs.\cite{Martin:2004,Martin:2005&2006,
Groeneboom:2008,Ren:2006}.

In the standard inflationary scenario, initial conditions for the
inflaton field are imposed in the infinite past with an infinitely
short wavelength when the effect of the inflationary horizon and the
expansion of the universe can be ignored. The spacetime is
essentially Minkowskian and there is a unique vacuum, the
Bunch-Davis vacuum, for the inflaton field. To encode the new
physics near the Planck scale, a simple modification in the standard
scenario has been focused on different choices of the vacuum
\cite{Goldstein:2003,Danielsson:2002b,Brandenberger:2002,WangX:2003,
Burgess:2003,Chernikov:1968,Mottola:1985,Allen:1985,Floreanini:1987,
Bousso:2002,Spradlin:2002}. It has been shown that based on some
choices of vacua  discussions of many trans-Planckian effects can be
carried out and qualitative correction due to Planck scale can be
obtained without detailed knowledge of trans-Planckian physics.
However, all these vacua chosen to discuss the inflation are of zero
temperature.

Inflation started just shortly after the big bang when the universe
was extremely hot at that moment. Before the inflation it would be
reasonable to consider that the universe was in thermal equilibrium.
After the inflation, the thermal equilibrium can be restored.  In
the process of the inflation, the inflaton coupled extremely weak to
thermal fields so that the inflaton itself is out of the thermal
equilibrium during inflation. However, considering that the
inflation process is extremely short, one can suppose that this
process can be described with the comoving temperature
$\tilde{T}=Ta$ to relate the initial and final states
\cite{Guth:1985,Bhattacharya:2006}, where $a$ is the scale factor,
and use the thermal ground state for the inflaton. In our work we
will use this assumption and employ the thermo field dynamics to
discuss the problem in detail.

In thermo field dynamics \cite{Umezawa:1982}, suppose the
temperature of the form $T=1/\beta$, the Bogoliubov transformations
of the usual boson operators
\begin{eqnarray}
\hat{a}_k(\beta)&=&\cosh\theta_k\hat{a}_k-\sinh\theta_{-k}\hat{\tilde{a}}^\dagger_{-k};\nonumber\\
\hat{\tilde{a}}_k(\beta)&=&\cosh\theta_{-k}\hat{\tilde{a}}_{-k}-\sinh\theta_k\hat{a}^\dagger_k,
\end{eqnarray}
can be reversed into
\begin{eqnarray}
\hat{a}_k&=&\cosh\theta_k\hat{a}_k(\beta)+\sinh\theta_k\hat{\tilde{a}}^\dagger_k(\beta);\nonumber\\
\hat{\tilde{a}}_k&=&\cosh\theta_{-k}\hat{\tilde{a}}_{-k}(\beta)+\sinh\theta_{-k}\hat{a}^\dagger_{-k}(\beta),
\end{eqnarray}
where $\cosh\theta_k=\frac{1}{\sqrt{1-e^{-\beta\omega(k)}}}$,
$\sinh\theta_k=\frac{e^{-\beta\omega(k)}}{\sqrt{1-e^{-\beta\omega(k)}}}$
and spectrum $\omega(k)=\sqrt{p^2+m^2}=\sqrt{k^2/a^2+m^2}$.
Obviously $\cosh\theta_k=\cosh\theta_{-k}$ and
$\sinh\theta_k=\sinh\theta_{-k}$, then in the following we will
simply use $\theta_k$.

In above definition we introduced the tilde operators
$\hat{\tilde{a}}_k$ and $\hat{\tilde{a}}^\dagger_k$ in the
duplicated space in additional to the normal operators.
Correspondingly, the state space should also be doubled, which means
that we will not only have $|n\rangle$, but also
$|\widetilde{n}\rangle$ ($n=0,1,2,\cdots$). The tilde operators
operate on tilde space vector $|\widetilde{n}\rangle$ while the
normal operators operate on the normal state $|n\rangle$. Thermal
operators satisfy the commuting relations,
$[\hat{a}_k(\beta),\hat{a}^\dagger_k(\beta)]=1$,
$[\hat{\tilde{a}}_k(\beta),\hat{\tilde{a}}^\dagger_k(\beta)]=1$,
$[\hat{a}_k(\beta),\hat{\tilde{a}}_k(\beta)]=0$, which can be
verified directly from the commutators of the operators $\hat{a}_k$
and $\hat{\tilde{a}}_k$.

Now we express the thermal vacuum as $|\beta0\rangle$. From the
thermo field dynamics, we learn that thermal vacuum state of a boson
field is the infinite linear composition of the state
$|n,\widetilde{n}\rangle$:
\begin{equation}
|\beta0\rangle=\sqrt{1-e^{-\beta\omega}}\sum_{n=0}^{\infty}e^{-n\beta\omega/2}|n,\widetilde{n}\rangle,
\end{equation}
which satisfies
\begin{equation}
\hat{a}_k(\beta)|\beta0\rangle=\hat{\tilde{a}}(\beta)|\beta0\rangle=0;\qquad
\langle\beta0|\hat{a}^\dagger_k(\beta)=\langle\beta0|\hat{\tilde{a}}^\dagger(\beta)=0.
\end{equation}
With these formalism, we are in a position to examine the influence
of the thermal vacuum on the power spectrum of the inflation.

In zero temperature case, given a statefunction $\chi_k$ solved from
field equation in momentum space, we can build up the field operator
as
\begin{equation}
\hat{\chi}=\sum_k(\chi_k\hat{a}_k+\chi^*_k\hat{a}^\dagger_{-k}).
\end{equation}
In the scalar perturbation, the power spectrum is defined as
\cite{Liddle:2000}
\begin{eqnarray}
P_k=\left(\frac{H}{\dot{\chi}}\right)^2\frac{k^3}{2\pi^2}|\chi_k|^2,
\end{eqnarray}
where the last factor arises from the expectation value of the field
operator in the vacuum state,
\begin{equation}
\langle0|\hat{\chi}^\dagger\hat{\chi}|0\rangle=\sum_k|\chi_k|^2.
\end{equation}

In the thermal vacuum, the field operator $\hat{\chi}$ can be
expressed as
\begin{eqnarray}
\hat{\chi}&=&\sum_k\{\chi_k(\eta)[\cosh\theta_k\hat{a}_k(\beta)+\sinh\theta_k
\hat{\tilde{a}}^\dagger_k(\beta)]+\chi^*_k(\eta)[\cosh\theta_k\hat{a}^\dagger_{-k}(\beta)
+\sinh\theta_k\hat{\tilde{a}}_{-k}(\beta)]\},\nonumber\\
&=&\sum_k\{\cosh\theta_k[\chi_k(\eta)\hat{a}_k(\beta)+\chi^*_k(\eta)
\hat{a}^\dagger_{-k}(\beta)]+\sinh\theta_k[\chi^*_k(\eta)\hat{\tilde{a}}_{-k}(\beta)
+\chi_k(\eta)\hat{\tilde{a}}^\dagger_k(\beta)]\}
\end{eqnarray}
by substituting Eq.(2) into Eq.(5). We see now the field space is
composed of two parts, the normal space and the duplicate space,
with weights $\cosh\theta_k$ and $\sinh\theta_k$, respectively.

The particle number can be calculated as
\begin{eqnarray}
n_k&=&\langle\beta0|\hat{a}^\dagger_k\hat{a}_k|\beta0\rangle\nonumber\\
&=&(1-e^{-\beta\omega})\sum_ne^{-n\beta\omega}\langle
n,\widetilde{n}|\hat{a}^\dagger_k\hat{a}_k|n,\widetilde{n}\rangle\nonumber\\
&=&(1-e^{-\beta\omega})\sum_ne^{-n\beta\omega}\langle
n,\widetilde{n}|\hat{a}^\dagger_k\sqrt{n}|n-1,\widetilde{n}\rangle
\nonumber\\
&=&(1-e^{-\beta\omega})\sum_ne^{-n\beta\omega}n\nonumber\\
&=&\frac{1}{e^{\beta\omega}-1},
\end{eqnarray}
which is just the distribution of thermal equilibrium states.

The power spectrum of perturbation in the thermal vacuum becomes
\begin{eqnarray}
P_k(\beta)&=&\left(\frac{H}{\dot{\chi}}\right)^2\frac{k^3}{2\pi^2}
\langle\beta0|\hat{\chi}_k^\dagger\hat{\chi}_k|\beta0\rangle\nonumber\\
&=&\left(\frac{H}{\dot{\chi}}\right)^2\frac{k^3}{2\pi^2}
\left(\{\langle\beta0|\cosh\theta_k^2[\chi_k^*(\eta)\hat{a}^\dagger_k(\beta)+\chi_k(\eta)
\hat{a}_{-k}(\beta)][\chi_k(\eta)\hat{a}_k(\beta)+\chi^*_k(\eta)
\hat{a}^\dagger_{-k}(\beta)]|\beta0\rangle\}\right.\nonumber\\
&&\left.+\{\langle\beta0|\sinh\theta_k^2[\chi_k(\eta)\hat{\tilde{a}}^\dagger_{-k}(\beta)
+\chi_k(\eta)\hat{\tilde{a}}_k(\beta)][\chi_k^*(\eta)\hat{\tilde{a}}_{-k}(\beta)
+\chi_k(\eta)\hat{\tilde{a}}^\dagger_k(\beta)]|\beta0\rangle\}\right)\nonumber\\
&=&\left(\frac{H}{\dot{\chi}}\right)^2\frac{k^3}{2\pi^2}|\chi_k\cosh\theta_k|^2
+\left(\frac{H}{\dot{\chi}}\right)^2\frac{k^3}{2\pi^2}|\chi_k\sinh\theta_k|^2\nonumber\\
&=&P_k(\cosh^2\theta_k+\sinh^2\theta_k)=P_k\coth\frac{\beta\omega}{2},
\end{eqnarray}
where $P_k$ is the power spectrum obtained by considering the zero
temperature vacuum. For the massless scalar field, $\omega=p=k/a$.
The temperature effect appears in the factor $\coth\frac{k}{2aT}$,
which shows that the power spectrum gets modified due to the thermal
effect. When $T\rightarrow0$, Eq.(10) reduces to the usual result of
Eq.(6). In the above derivation, we use the finite temperature field
theory which clearly shows the temperature effect in the vacuum.
This approach is general and does not depend on the choice of the
vacuum.

Adding the thermal effect in the adiabatic vacuum, we can
generalize the power spectrum in \cite{Danielsson:2002a} by
including the thermal factor as
\begin{equation}
P_k(\beta)=P_k\coth\frac{\beta\omega}{2}=\left(\frac{H}{\dot{\chi}}\right)^2\left(\frac{H}{2\pi}\right)^2
\left[1-\frac{H}{\Lambda}\sin\left(\frac{2\Lambda}{H}\right)\right]\coth\frac{k}{2\tilde{T}},
\end{equation}
where $\tilde{T}=aT$ is the comoving temperature and $\Lambda$ is
the trans-Planckian energy level. The second term in the bracket
represents the effect brought by the trans-Planckian physics. At
zero temperature the modulation of the power spectrum of primordial
density fluctuation predicted in the trans-Planckian model has been
studied by considering the change of the Hubble parameter and
slow-roll condition \cite{Bergstrom:2002}
\begin{equation}
\varepsilon\equiv2M^2_{PL}\left(\frac{1}{H(\chi)}\frac{dH(\chi)}{d\chi}\right)^2,
\end{equation}
where $M_{PL}^{-2}=8\pi G$ is the reduced Planck mass. In
Ref.\cite{Bergstrom:2002}, adopting the scale parameter
$\gamma=\Lambda/M_{PL}$ and $H/\Lambda=\xi(k/k_0)^{-\varepsilon}$,
the trans-Planckian power spectrum is expressed into
\begin{equation}
P_k=\left(\frac{H}{\dot{\chi}}\right)^2\left(\frac{H}{2\pi}\right)^2
\left\{1-\xi\left(\frac{k}{k_0}\right)^{-\varepsilon}\sin\left[\frac{2}{\xi}
\left(\frac{k}{k_0}\right)^\varepsilon\right]\right\},
\end{equation}
where $\xi\approx4\times10^{-4}\sqrt{\varepsilon}/\gamma$ and the
pivot scale $k_0=0.05$ Mpc$^{-1}\approx213.8H_0$, based on the value
$H_0=70.1$ km s$^{-1}$ Mpc$^{-1}$ from five-year WMAP result
\cite{Hinshaw:2008}. The power spectra $P(k)$ for different values
of $\gamma$ and $\varepsilon$ were shown in
Ref.\cite{Bergstrom:2002}. In Fig.1 we illustrate their result of
zero temperature case in the solid line by taking
$\varepsilon=\gamma=0.01$.

\begin{figure}[tbp]
\includegraphics[totalheight=12cm, width=16cm]{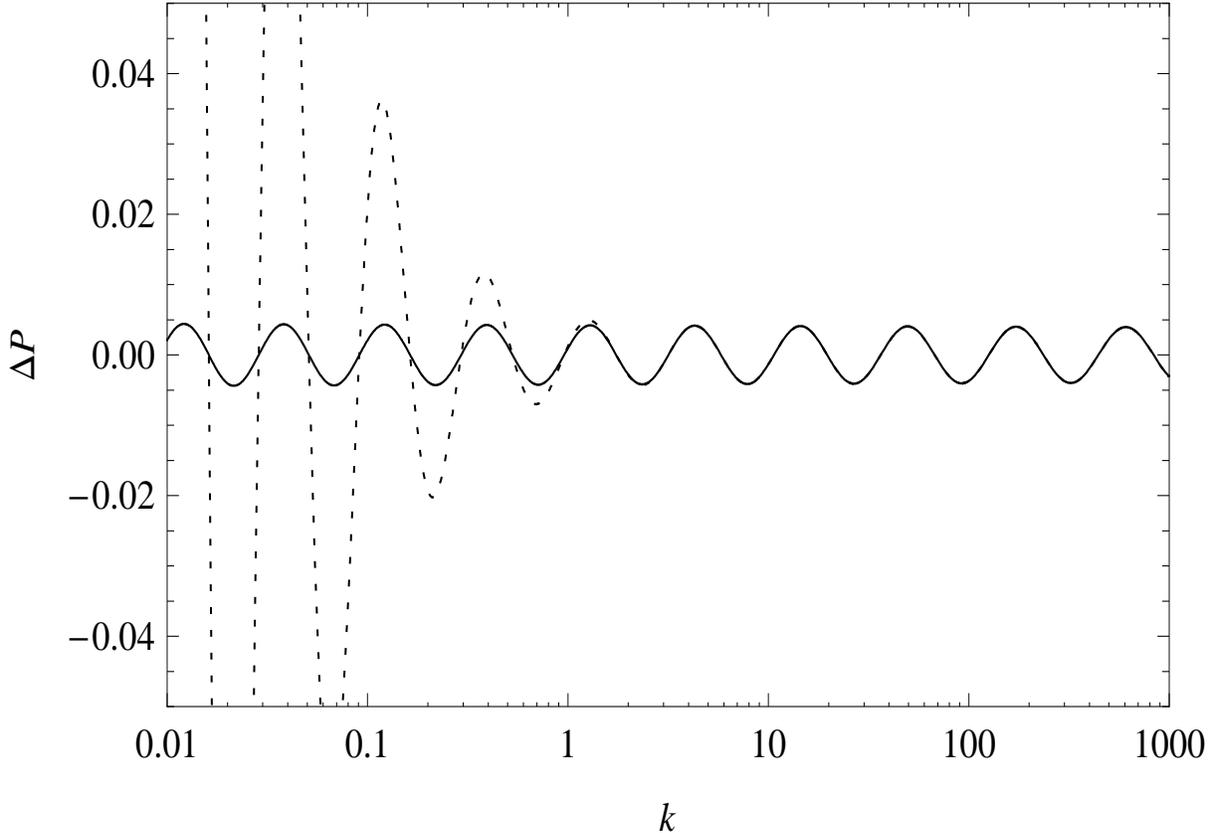}
\caption{The modulation of the power spectrum of primordial density
fluctuations predicted by the trans-Planckian and the thermal
effects. The solid line is for zero temperature while the dashed
line shows the thermal influence. The parameters are
$\varepsilon=0.01$ and $\gamma=0.01$.} \label{fig1}
\end{figure}

\begin{figure}[tbp]
\includegraphics[totalheight=12cm, width=16cm]{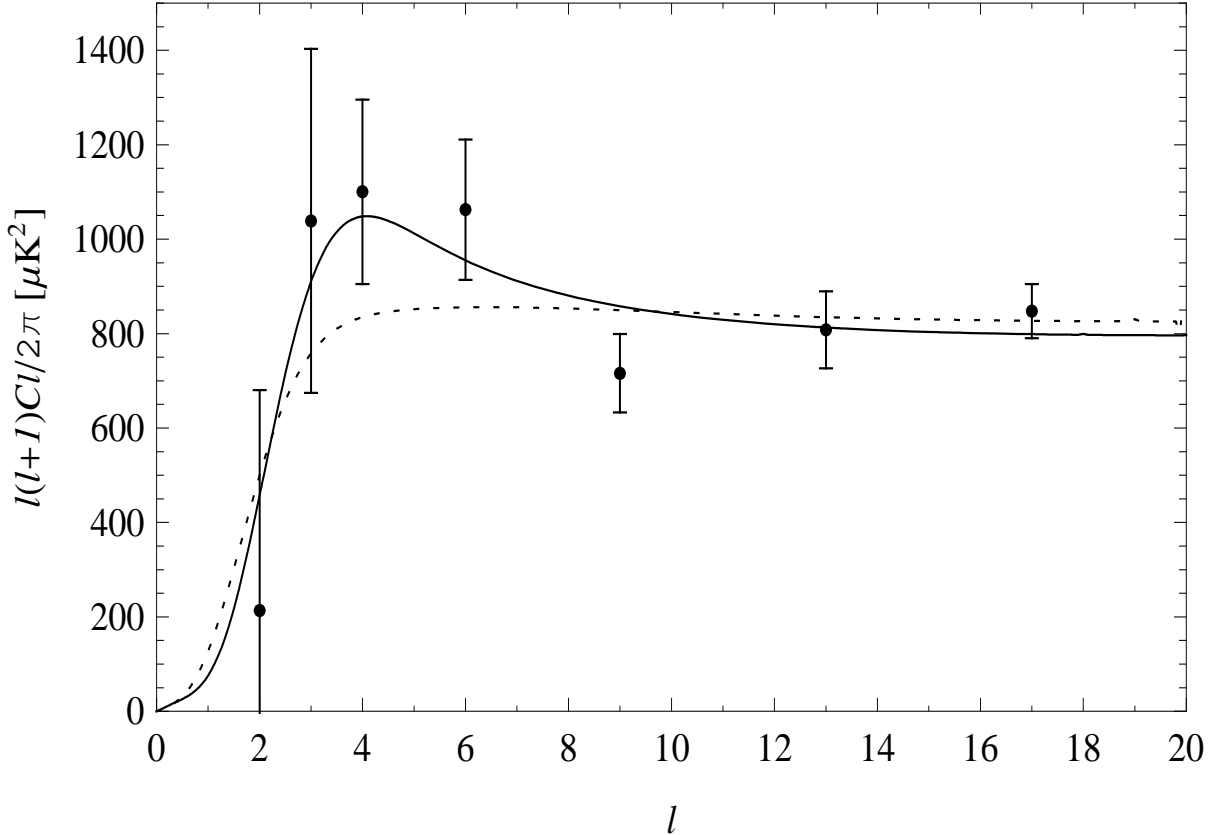}
\caption{The angular power spectrum at low $l$ and its comparison
with WMAP data. The dashed line is for zero temperature case, while
the solid line shows the thermal effect. The best fitting of solid
line is made with $c=2.3$ and $\tilde{T}=0.9H_0$, while for the
dashed line $c=3.1$.} \label{fig2}
\end{figure}

Considering the thermal effect with additional term
$\coth\frac{k}{2\tilde{T}}$, we have the power spectrum
\begin{equation}
P_k(\beta)=\left(\frac{H}{\dot{\chi}}\right)^2\left(\frac{H}{2\pi}\right)^2
\left\{1-\xi\left(\frac{k}{k_0}\right)^{-\varepsilon}\sin\left[\frac{2}{\xi}
\left(\frac{k}{k_0}\right)^\varepsilon\right]\right\}\coth\frac{k}{2\tilde{T}}.
\end{equation}
The temperature influence on the power spectrum $P_k$ is shown by
the dashed line in Fig.1. We took $\tilde{T}=H_0$ in our plot and
noticed that different values of $\tilde{T}$ will not change the
qualitative behavior of the dashed line. The thermal influence
becomes significant especially at small $k$, while its influence can
be neglected for big enough $k$. In other words the change in the
spectrum due to the thermal effect is large at large angles and
small at small angles. This brings an interesting possibility that
the thermal effect might influence the CMB at small $l$.

When $k\rightarrow0$, because of the factor
$\coth\frac{k}{2\tilde{T}}$, we will meet the divergence problem
in the power spectrum, which is unphysical. To tackle this problem
we resort to the idea of the holographic cutoff employed in
\cite{Enqvist:2004,Shen:2005}. Relating the ultraviolet and
infrared cutoff \cite{Cohen:1999}, the quantum zero-point energy
density $\rho_\Lambda$ which can be related to the cosmological
constant or dark energy density in a flat universe can be
expressed by
\begin{equation}
\rho_\Lambda=3c^2M_{PL}^2L^{-2},
\end{equation}
where $L$ is the infrared cutoff, and $c$ is a constant parameter.
From this relation, we can write
$L=c/(\sqrt{\Omega_{\Lambda0}}H_0)$, where $\Omega_{\Lambda0}$ is
the present value of the dark energy ratio to the critical density
$\Omega_\Lambda=\rho_\Lambda/\rho_{cr}=\rho_\Lambda/(3M_{PL}^2H^2)$.
Consequently this introduces the cutoff at the wave number
\cite{Shen:2005}
\begin{equation}
k_c=\frac{\pi}{c}\sqrt{\Omega_{\Lambda0}}H_0.
\end{equation}
This holographic cutoff can help to remove the divergency brought by
$k=0$.

Now we move on to examine the thermal effect on the CMB at small $l$
by comparing with the WMAP data. The CMB angular spectrum is
expressed as \cite{Giovannini:2005}
\begin{equation}
C_l=\frac{4\pi}{9}\int_{k_c}^\infty\frac{dk}{k}j_l^2(k\Delta\tau)P_k,
\end{equation}
where $j_l$ is the spherical Bessel function and $\Delta\tau$ is the
comoving distance to the last scattering surface,
\begin{equation}
\Delta\tau=\int_0^{z_0}\frac{dz}{H(z)},
\end{equation}
and $z_0$ is the redshift of decoupling usually taken as $z_0=1100$.
We choose the Hubble parameter
\begin{equation}
H(z)=H_0\sqrt{(1-\Omega_{\Lambda0})(1+z)^3+\Omega_{\Lambda0}(1+z)^{3(1+w)}},
\end{equation}
where $w=p/\rho$ is the equation of state of dark energy and is
assumed as a constant for simplicity. In our numerical calculation,
we will set $\Omega_{\Lambda0}=0.721$ and $w=-0.972$
\cite{Hinshaw:2008}. For the zero temperature case, it has been
shown that the relation between ultraviolet and infrared cutoff can
help to explain the small $l$ CMB suppression \cite{Shen:2005},
which is exhibited in the dashed curve in Fig.2. However, compared
with the WMAP data, it cannot explain well the wriggle observed at
$l=3,4,6$.

Notice that the thermal effect may play the role at large angle to
modify the power spectrum, we now turn to the temperature dependent
spectrum in Eq.(14). To do the data fitting with WMAP result at
small $l$, we have parameters $\tilde{T}$, $c$, $\varepsilon$ and
$\xi$ or $\gamma$ now. We observed that in the reasonable range of
slow-roll condition, values of $\varepsilon$ and $\gamma$ do not
significantly change the total power spectrum, thus we fix them to
be $\varepsilon=0.03$ and $\gamma=0.003$ to ensure observable
trans-Planckian effect to be found in CMB \cite{Bergstrom:2002}. Now
we have two parameters $c$ and $\tilde{T}$ to be determined.
Calculating the angular power spectrum $l(l+1)C_l$ and comparing
with WMAP data by doing $\chi^2$ fitting for $0<l\leq20$,
\begin{equation}
\chi^2=\sum_i\frac{[l(l+1)C_l^i|_{theory}-l(l+1)C_l^i|_{data}]^2}
{(\sigma^i)^2}
\end{equation}
where $\sigma^i$ is the observational error of each data, we
present the best fitting result in solid line in Fig.2, where
$c=2.5$ and $\tilde{T}=0.9H_0\approx1.3\times10^{-33}$ eV
$\approx1.6\times10^{-29}$ K. This result keeps almost the same
when we shift $\varepsilon$ and $\gamma$ in the reasonable range
of the slow-roll inflation condition. It can be seen from Fig.2
that the thermal effect can well explain the WMAP data at small
$l$, especially the wriggle at $l=3,4,6$.

In summary, we have employed the thermo field dynamics to
investigate the thermal vacuum effect on the power spectrum of the
inflation. Comparing with the spectrum of the zero temperature case,
we have observed that the thermal effect plays the role essentially
in the low multipoles or large length scale. Resorting to the idea
of holographic cutoff, we have removed the divergence problem when
$k=0$. Comparing with the WMAP data at small $l$, we have found that
the thermal effect explains well the data at small $l$ than the zero
temperature case. When $l=3,4,6$, the spectrum got enhanced due to
the temperature effect. This suggests that the thermal effect should
be considered in studying the inflation, especially when we want to
study the CMB anisotropy at low multipoles. The temperature was very
high at the beginning of the inflation, so it actually influences
the fluctuation spectrum which escaped earlier from the horizon and
reentered later so that the large scale (small $l$) CMB spectrum got
corrected. When the inflation started, the temperature dropped fast,
and the temperature effect has less influence on the fluctuation
spectrum. These fluctuations left the horizon later and reentered
earlier which affect the small scale CMB spectrum. Since the
temperature effect is low, most CMB spectrum ($l>10$) has little
difference compared with the zero temperature case. This actually
gives the reason why we see the temperature correction is important
in the small $l$ CMB spectrum, which is consistent with the
observation, while for big $l$, the temperature effect is
negligible.

\section*{Acknowledgments}
This work was partially supported by the NNSF of China, Shanghai
Education Commission, Shanghai Science and Technology Commission. We
would like to acknowledge helpful discussions with R.G.Cai and
Y.G.Gong. S. Yin's work was also partially supported by the graduate
renovation foundation of Fudan university.

\end{document}